\documentstyle[11pt]{article}
\author{H. Liebl\\
Department of Physics, Tufts University, Medford, MA 02155
\and
Gary R. Goldstein\footnotemark\\
Center for Theoretical Physics, Massachusetts Institute of Technology\\
Cambridge, MA 02139}
\footnotetext{Permanent address: Department of Physics, Tufts University,
Medford, MA 02155}
\title{Electromagnetic Polarizabilities and Charge Radii of the Nucleons
in the Diquark-model}
\date{Sept.1994 }

\begin{document}
\maketitle
\begin{abstract}
The diquark model is used to calculate the electromagnetic polarizabilities
and charge radii of the nucleons for three different potentials. Making the
scalar diquark lower in mass introduces a mixing angle $\theta $ between the
$\left| 56\right\rangle $ and $\left| 70\right\rangle $ states ,which allows
an improvement in value of all 6 properties. Generalizing the Gamov-Teller
matrix and the magnetic moment operator to the diquark model gives
constraints on this mixing. We obtain for the Richardson potential $\theta
=23.2^{\circ },$ $\overline{\alpha }_p=7.9_{-0.9}^{+1.0}\times 10^{-4}fm^3,$
$\overline{\alpha }_n=7.7_{-0.6}^{+0.3}\times 10^{-4}fm^3,$ $\overline{\beta
}_p=5.4_{-0.4}^{+1.6}\times 10^{-4}fm^3,$ $\overline{\beta }%
_n=6.7_{-0.7}^{+1.3}\times 10^{-4}fm^3,$ $\left\langle r^2\right\rangle
_p=0.37_{-0.03}^{+0.02}fm^2,$ $\left\langle r^2\right\rangle
_n=-0.07_{-0.02}^{+0.03}fm^2.$ Additional pion cloud contributions could
improve on all six results.

\newpage
\end{abstract}

The electric and magnetic polarizabilities, labeled $\overline{\alpha }$ and
$\overline{\beta }$ ,respectively, have been measured recently\cite
{schmiedmayer}$^{,}$\cite{federspiel}, yielding the results $\overline{%
\alpha }_p=\left( 10.9\pm 2.2\pm 1.3\right) \times 10^{-4}fm^3,$ $\overline{%
\beta }_p=\left( 3.3\mp 2.2\mp 1.3\right) \times 10^{-4}fm^3,$ $\overline{%
\alpha }_n=\left( 12.3\pm 1.5\pm 2.0\right) \times 10^{-4}fm^3$ and $
\overline{\beta }_n=\left( 3.5\mp 1.5\mp 2.0\right) \times 10^{-4}fm^3.$ The
experimental values are obtained by fitting the Compton scattering data to
relations obtained from low energy theorems\cite{friar}. In addition to the
polarizabilities we will be concerned here with the charge radii $%
\left\langle r^2\right\rangle =\sum e_ix_i^2,$ whose values were measured to
be\cite{koestner} $\left\langle r^2\right\rangle _p=0.708fm^2$ and $%
\left\langle r^2\right\rangle _n=-0.113\pm 0.003fm^2.$ Theoretical studies,
using the 3 symmetric quark model\cite{holstein}, Skyrme models\cite
{nymanchemtop}, MIT-bag models\cite{heckingschaefer} or nonlinear meson
theories\cite{scoccola} have not obtained satisfactory results for all these
quantities. A recent paper using chiral perturbation theory\cite{bernard} gives
good agreement for $\overline{\alpha }$, but predicts negative values for
the magnetic polarizabilities.

In this paper we will use another approach, the diquark model. This model,
which provides improved agreement with a wide range of data\cite{anselmino},
reduces the three body problem of the quark picture to a two body problem
and results, therefore, in a considerable simplification of the calculation of
baryonic properties. To review the salient features, the proton and neutron
wavefunctions have spin-flavour forms
\begin{equation}
\label{mixp}|p,\uparrow \rangle =\frac 1{\sqrt{1+a^2}}\left| \left( \sqrt{%
\frac 23}S_{uu}d-\sqrt{\frac 13}S_{ud}u\right) \left( \sqrt{\frac 23}%
|1,\downarrow \rangle -\sqrt{\frac 13}|0,\uparrow \rangle \right)
+at_{ud}u|0,\uparrow \rangle \right\rangle
\end{equation}
\begin{equation}
\label{mixn}|n,\uparrow \rangle =\frac 1{\sqrt{1+a^2}}\left| \left( \sqrt{%
\frac 23}S_{dd}u-\sqrt{\frac 13}S_{ud}d\right) \left( \sqrt{\frac 23}%
|1,\downarrow \rangle -\sqrt{\frac 13}|0,\uparrow \rangle \right)
+at_{du}d|0,\uparrow \rangle \right\rangle
\end{equation}
where S and t are two different diquark states (S for sextet and t for
triplet of SU(3) flavour) with spin S=1 and S=0, respectively. The parameter
a=1 for the fully SU(6) symmetric scheme. The spin interaction between the
two quarks, which form the diquark, is assumed to yield a mass splitting
between the S- and the t-state and breaks SU(6) symmetry. From the QCD
hyperfine interaction $\bigtriangleup m\propto \stackrel{\rightharpoonup }{S}%
_1\cdot \stackrel{\rightharpoonup }{S}_2$ it follows that the S-state is
heavier than the t-diquark. Hence a will differ from 1, as we discuss below.
In addition there will be another mass splitting $\bigtriangleup m_2$ coming
from the spin interaction between the quark and the diquark, which increases
the spin $\frac 32$ mass relative to the spin $\frac 12$ state.

To determine the spatial dependence of the wavefunctions we use three
different forms of scalar potentials, a logarithmic one $V_L\left( r\right)
=C\cdot Log\left( \frac r{r_o}\right) ,$ a superpostion of linear and
Coulomb-potential $V_{CL}\left( r\right) =-\frac{4\alpha _s}{3r}+Cr$ and the
Richardson potential\cite{richardson} $V_R,$ which interpolates the
Coulombic potential for small r and the linear potential for large r.

The expressions for the polarizabilities, derived for the nuclei of atoms%
\cite{friar} can be modified for the nucleons, yielding
\begin{equation}
\label{1}\overline{\alpha }=\frac{\alpha \sum e_i}{3M}\left\langle \sum
e_ix_i^2\right\rangle +\frac{2\alpha }3\sum\limits_n\frac{\left|
\left\langle 0\right| \stackrel{\rightharpoonup }{D}\left| n\right\rangle
\right| ^2}{E_n-E_o}
\end{equation}
\begin{equation}
\label{2}\overline{\beta }=2\alpha \sum\limits_n\frac{\left| \left\langle
0\right| \sum \mu _i^z\left| n\right\rangle \right| ^2}{E_n-E_o}-\frac
\alpha 6\left\langle \sum \frac{e_i^2x_i^2}{m_i}\right\rangle -\frac{%
\alpha \left\langle D^2\right\rangle }{2M}
\end{equation}
where $\alpha $ is the fine structure constant, M the mass of the nucleon, m$%
_i$ the mass of its constituents, $\stackrel{\rightharpoonup }{D}$ the
electric dipole operator and $\mu _i$ the magnetic moment operator.

The nonrelativistic wavefunctions for all three potentials were solved
numerically. To get the parameters of the different potentials we fitted the
eigenenergies to different baryon states, the N(938), N(1710), $\Delta $%
(1232) and $\Delta $(1600). In the diquark picture\cite{haire} they
correspond to the (56,0), (56,0)$^{**}$ (56,0) and (56,0)$^{*}$,
respectively, where the $\Delta $ wavefunction
$$
\left| \Delta ^{+},\uparrow \right\rangle =\left| \left( \sqrt{\frac 13}%
S_{uu}d+\sqrt{\frac 23}S_{ud}u\right) \left( \sqrt{\frac 13}\left|
1,\uparrow \right\rangle +\sqrt{\frac 23}\left| 0,\uparrow \right\rangle
\right) \right\rangle
$$
has to be used for the two latter states.The dipole operator requires that
the intermediate states of the electric polarizabilities eq.(\ref{1}) are
all L=1 states. Therefore we have to take (56,1) and (70,1) states as well
as their excitations. The $\left| 56\right\rangle $ states have the same
form as eqs.(\ref{mixp}) (\ref{mixn}) and the $\left| 70\right\rangle $
states are proportional to\cite{goldstein}%
$$
\left| 70^{+}\right\rangle =\left| \left( \sqrt{\frac 13}S_{uu}d-\sqrt{%
\frac 16}S_{ud}u\right) \left( \sqrt{\frac 23}|1,\downarrow \rangle -\sqrt{%
\frac 13}|0,\uparrow \rangle \right) -\sqrt{\frac 12}t_{ud}u|0,\uparrow
\rangle \right\rangle
$$

Excited states soon give a negligible contribution to $\overline{\alpha }$,
since the denominators $E_n-E_0$ increase as the numerator decreases. It is
sufficient to take the 2p and 3p wavefunctions into account. As we will see
below it is necessary to introduce a mixing between the $\left|
56\right\rangle $ and the $\left| 70\right\rangle $ states. To see the
essential features of a ''symmetric diquark model'' in what follows we would
treat the nucleons as if they were pure $\left| 56\right\rangle $ states
(i.e. set $a=1$ in the wavefunctions). Defining $C_{1t,s}\equiv \frac
m{m+m_{t,s}}$ and $C_{2t,s}\equiv \frac{m_{t,s}}{m+m_{t,s}}$ we get for the
electric polarizabilities

\newcommand{\D}{\displaystyle}
\normalsize

\begin{equation}
\label{alphp}
\begin{array}[t]{lll}
\overline{\alpha }_p & = & {\D \frac{\alpha}{9M}}{\D\frac 1 {\left(
1+a^2\right)
}}\left( \left\langle r_t^2\right\rangle a^2\left( C_{1t}^2+2C_{2t}^2\right)
+3\left\langle r_s^2\right\rangle C_{1s}^2\right) \\
& + & \D\sum\limits_{n=2,3}
\frac{2\alpha }{3\left( 1+a^2\right) ^2}\left( \frac{a^2\left\langle
56,np\right| r_t\left| 1s\right\rangle }{\sqrt{E_{np,t}-E_{1s,t}}}\left(
\frac 13C_{1t}-\frac 23C_{2t}\right) +\frac{\left\langle 56,np\right|
r_s\left| 1s\right\rangle }{\sqrt{E_{np,s}-E_{1s,s}}}C_{2s}\right) ^2 \\  &
+ & \D\sum\limits_{n=2,3}\frac{2\alpha }{3\left( 1+a^2\right) ^2}\left(
\frac{-a^2\left\langle 70,np\right| r_t\left| 1s\right\rangle }{\sqrt{%
E_{np,t}-E_{1s,t}}}\left( \frac 13C_{1t}-\frac 23C_{2t}\right) +\frac{%
\left\langle 70,np\right| r_s\left| 1s\right\rangle }
{\sqrt{E_{np,s}-E_{1s,s}}}%
C_{2s}\right) ^2
\end{array}
\end{equation}

\begin{equation}
\label{alphn}
\begin{array}[t]{lll}
\overline{\alpha }_n & = & \D\sum\limits_{n=2,3}
\frac{2\alpha }{27\left( 1+a^2\right) ^2}\left( \frac{a^2\left\langle
56,np\right| r_t\left| 1s\right\rangle }{\sqrt{E_{np,t}-E_{1s,t}}}
 -\frac{\left\langle 56,np\right|
r_s\left| 1s\right\rangle }{\sqrt{E_{np,s}-E_{1s,s}}}\right) ^2 \\  &
+ & \D\sum\limits_{n=2,3}\frac{2\alpha }{9\left( 1+a^2\right) ^2}\left(
\frac{-a^2\left\langle 70,np\right| r_t\left| 1s\right\rangle }{\sqrt{%
E_{np,t}-E_{1s,t}}} -\frac{%
\left\langle 70,np\right| r_s\left| 1s\right\rangle }
{\sqrt{E_{np,s}-E_{1s,s}}}%
\right) ^2
\end{array}
\end{equation}

where $r_{s,t}$ are the coordinates of the reduced mass.

For the magnetic polarizabilities the first term of eq. (\ref{2}) needs some
modification. The magnetic moment operator has to be generalized for the
quark-diquark states to
\begin{equation}
\label{muz}\mu _z=g\frac{e_d}{2m_d}S^z+\frac{e_3}{2m_3}\sigma ^z
\end{equation}
However this alone leads to problems when calculating the ratio of the
magnetic moments $\frac{\mu _p}{\mu _n}$. Only a value of g=0 would
reproduce the successful predictions of the symmetric quark model of $-1.5$
(compared to the experimental value of $-1.46)$. Analyzing the difference
between the diquark picture and the latter reveals the need for a
contribution coming from $\left\langle S_{ud}\right| \mu _z\left|
t_{ud}\right\rangle ,$ which is non-zero for a composite diquark. Therefore
we add $\left\langle S_{ud}\right| \frac{e_1}{2m_1}\sigma ^z+\frac{e_2}{2m_2}%
\sigma ^z\left| t_{ud}\right\rangle $ to eq.(\ref{muz}). If we include this
term we get exactly the quark result again, in the limit g$\rightarrow 2$
and $m_d\rightarrow 2m.$ This shows how important it is not to see the
diquark naively as a single point particle but rather as a composite of two
quarks. To continue, it is sufficient to take the $\Delta $ as the dominant
intermediate state\cite{heckingnyman} in eq.(\ref{2}) and neglect the higher
mass states. The second and third terms of eq.(\ref{2}) are straightforward
to calculate and so the magnetic polarizabilities become

\begin{equation}
\label{betap}
\begin{array}{lll}
\overline{\beta }_P & = &\D \frac{2\alpha }{81\left( M_\Delta -M_p\right)
\left( 1+a^2\right) }\left( \frac{2+3a}m+\frac g{m_s}\right) ^2 \\  & - &
\D \frac \alpha {54\left( 1+a^2\right) }\left( a^2\left\langle
r_t^2\right\rangle \left(
\frac{C_{1t}^2}{m_t}+\frac{4C_{2t}^2}m\right) +\left\langle
r_s^2\right\rangle \left( \frac{11}{m_s}C_{1s}^2+\frac 2mC_{2s}^2\right)
\right)  \\  & - & \D \frac \alpha {54M\left( 1+a^2\right) }\left(
3a^2\left\langle r_t^2\right\rangle \left( C_{1t}-2C_{2t}\right)
^2+\left\langle r_s^2\right\rangle \left( 2\left( 4C_{1s}+C_{2s}\right)
^2+\left( C_{1s}-2C_{2s}\right) ^2\right) \right)
\end{array}
\end{equation}

\begin{equation}
\label{betan}
\begin{array}{lll}
\overline{\beta }_n & = & \D \frac{2\alpha }{81\left( M_\Delta -M_p\right)
\left( 1+a^2\right) }\left( \frac{2+3a}m+\frac g{m_s}\right) ^2 \\  & - &
\D \frac \alpha {54\left( 1+a^2\right) }\left( a^2\left\langle
r_t^2\right\rangle \left(
\frac{C_{1t}^2}{m_t}+\frac{C_{2t}^2}m\right) +3\left\langle
r_s^2\right\rangle \left( \frac{C_{1s}^2}{m_s}+\frac{C_{2s}^2}m\right)
\right)  \\  & - & \D \frac \alpha {54M\left( 1+a^2\right) }\left(
3a^2\left\langle r_t^2\right\rangle \left( C_{1t}+C_{2t}\right)
^2+\left\langle r_s^2\right\rangle \left( 8\left( C_{1s}+C_{2s}\right)
^2+\left( C_{1s}+C_{2s}\right) ^2\right) \right)
\end{array}
\end{equation}

The charge radii in the diquark picture are

\begin{equation}
\label{rp}
\begin{array}{ll}
\left\langle r_P^2\right\rangle & \D =\frac{a^2\left\langle r_t^2\right\rangle
}{3\left( 1+a^2\right) }\left( C_{1t}^2+2C_{2t}^2\right) + \frac{%
\left\langle r_s^2\right\rangle }{1+a^2}C_{1s}^2
\end{array}
\end{equation}
\begin{equation}
\label{rn}
\begin{array}{ll}
\left\langle r_N^2\right\rangle & = \D \frac 1{3\left( 1+a^2\right) }\left(
a^2\left\langle r_t^2\right\rangle \left( C_{1t}^2-C_{2t}^2\right)
+\left\langle r_s^2\right\rangle \left( C_{1s}^2-C_{2s}^2\right) \right)
\end{array}
\end{equation}

Using these formulas with $a=1$ we obtain the results shown in table 1.%
\begin{table}
  \begin{tabular}{|l|l|l|l|} \hline
     {\bf }  &  {\bf  Logarithmic}&{\bf Lin.-Coulomb.} &  {\bf  Richardson} \\
\hline

  $\overline{\alpha }_p$ $[fm^3]$  $\times 10^{-4}$ &     $4.1\pm 0.4$    &
$5.0\pm 0.4$    & $5.2\pm 0.5$  \\
  $\overline{\alpha }_n$ $[fm^3]$ $\times 10^{-4}$ &   $3.3\pm 0.1$
&$4.5\pm 0.1$   &   $4.4\pm 0.2$ \\
  $\overline{\beta }_p$ $[fm^3]$ $\times 10^{-4}$  &  $3.9\pm 0.2$        &
$7.8\pm 0.4$  &  $7.6\pm 0.2$     \\
  $\overline{\beta }_n$ $[fm^3]$ $\times 10^{-4}$  & $4.1\pm 0.2$         &
$8.4\pm 0.4$  &  $9.0\pm 0.4$  \\
  $\left\langle r^2\right\rangle _p$ $[fm^2]$ &  $0.20\pm 0.02$      & $0.25\pm
0.01$ & $0.25\pm 0.02$      \\
  $\left\langle r^2\right\rangle _n$ $[fm^2]$ &  $0.005\pm 0.004$ & $0.012\pm
0.006$&     $0.015\pm 0.006$ \\ \hline
  \end{tabular}
  \caption{Results for the polarizabilities and charge radii for different
forms of potentials with no mixing, a=1}
\end{table}
The values of the electric polarizabilities are too small, the magnetic
quantities too big and the charge radii too small. Note especially that the
neutron charge radius has the wrong sign and is very close to zero.

So far we have treated the nucleons as pure $\left| 56\right\rangle $
states. The situation turns out quite favourable if we introduce a mixing
angle $\theta $ between the $\left| 56\right\rangle $ and the $\left|
70\right\rangle $ states to form the nucleons
\begin{equation}
\label{angle}\left| N\right\rangle =\cos \theta \left| 56\right\rangle +\sin
\theta \left| 70\right\rangle
\end{equation}
which is equivalent to introducing a mixing parameter $a,$ for which $\cos
\theta =\frac{a+1}{\sqrt{2a^2+2}}.$ The QCD hyperfine splitting as well as
phenomenological studies\cite{frederiksson} suggest that the S=0 state has
lower mass and thus is a larger component of the nucleon. We will use values
$a\geq 1$ in the above equations. To get an upper constraint on $a$, we fit
our model to the ratio of the nucleon magnetic moments $\frac{\mu _p}{\mu _n}
$ as well as to the axial to vector current coupling constant ratio $\frac{%
g_a}{g_v}$.

Using the above magnetic moment operator for the diquark model and the
wavefunctions eqs.(\ref{mixp}),(\ref{mixn}) we obtain
\begin{equation}
\label{mupton} \D \frac{\mu _p}{\mu _n}=-3\frac{\D \left( \frac{a^2+a}m+\frac
g{m_s}\right) }{\D \left( \frac{3a^2+6a+1}{2m}+\frac g{m_s}\right) }
\end{equation}
The operator for $\frac{g_a}{g_v},$ also called the Gamov-Teller matrix\cite
{close} is
\begin{equation}
\label{gamov}\frac{\left\langle p,\uparrow \right| \sum_i\tau
_i^{+}S_i^Z\left| n,\uparrow \right\rangle }{\left\langle p,\uparrow \right|
\sum_i\tau _i^{+}\left| n,\uparrow \right\rangle }
\end{equation}
As we generalized the magnetic moment operator to the diquark model we will
also have to generalize the Gamov-Teller matrix. Analogously we start with%
$$
\sum\limits_i\tau _i^{+}S_i^z=\tau _3^{+}\sigma _3^z+hT_d^{+}S_d^z
$$
h can be factorized as $h=t*g$ and comparing the terms to the quark picture
shows that t$=\frac 1{\sqrt{2}}.$ As in the case of the magnetic moments we
have to add terms from the quark picture,%
$$
\left( \tau _1^{+}\sigma _1^z+\tau _2^{+}\sigma _2^z\right) \left|
t_{ud},0\right\rangle =\sqrt{2}\left| S_{uu},0\right\rangle
$$
$$
\left( \tau _1^{+}\sigma _1^z+\tau _2^{+}\sigma _2^z\right) \left|
S_{dd},0\right\rangle =\sqrt{2}\left| t_{ud},0\right\rangle
$$

The denominator of eq.(\ref{gamov}) remains $-1$ also for the case of mixing
and so we get
\begin{equation}
\label{gagv}\frac{g_a}{g_v}=-\left\langle p,\uparrow \right| \sum_i\tau
_i^{+}S_i^Z\left| n,\uparrow \right\rangle =\frac{1+12a+9a^2+4g}{9\left(
1+a^2\right) }
\end{equation}

Now we take the ratio of the magnetic moments fixed at its experimantal
value $\frac{\mu _p}{\mu _n}=-1.46$\cite{datagroup} and use eq.(\ref
{mupton}) to calculate $g$ as a function of $a$ and $\frac{m_s}m,$ which in
turn (eq.(\ref{gagv})) gives $\frac{g_a}{g_v}$ as another function of those
parameters. The dependence on $\frac{m_s}m$ is weak compared to the $a$%
-dependence. A value of $a=2.5\pm 0.1$ fits the experimental value of
$\frac{g_a}{g_v}%
=1.25$\cite{datagroup}. Since taking the relativistic nucleon wavefunctions
would have a lowering effect anyway on $\frac{g_a}{g_v}$\cite{bogoliubov} we
will take $a=2.5$ as the upper limit and take $\bigtriangleup a=-0.3$ for a
theoretical error calculation. In terms of the mixing angle of eq.(\ref
{angle}) $a=2.5$ means $\theta =23.2^{\circ }$ and therefore gives a small $%
\left| 70\right\rangle $ state contribution to the nucleons.

Figure 1 shows the results for our quantities as a function of the mixing
factor $a$ for the Richardson potential. The situation for the other
potentials looks qualitatively similar. We see immediately that a large
mixing factor improves the data for all 6 quantities and justifies our
approach.

Table 2 shows the results for all 3 forms of potentials, where the diquark
mass splitting was taken as $\bigtriangleup m=100\pm 50MeV.$%
\begin{table}
  \begin{tabular}{|l|l|l|l|} \hline
     {\bf }  &  {\bf  Logarithmic}&{\bf Lin.-Coulomb.} &  {\bf  Richardson} \\
\hline

  $\overline{\alpha }_p$ $[fm^3]$  $\times 10^{-4}$& $5.8_{-0.8}^{+0.4}$
&$8.4_{-1.5}^{+1.3}$ &  $7.9_{-0.9}^{+1.0}$   \\
  $\overline{\alpha }_n$ $[fm^3]$ $\times 10^{-4}$ & $5.2_{-0.5}^{+0.4}$ &
$8.4_{-0.6}^{+0.3}$ & $7.7_{-0.6}^{+0.3}$   \\
  $\overline{\beta }_p$ $[fm^3]$  $\times 10^{-4}$ & $2.0_{-0.1}^{+0.3}$  &
$6.5_{-0.6}^{+1.0}$  &  $5.4_{-0.4}^{+1.6}$     \\
  $\overline{\beta }_n$ $[fm^3]$  $\times 10^{-4}$ & $2.7_{-0.2}^{+0.7}$ &
$7.9_{-0.9}^{+2.0}$  &  $6.7_{-0.7}^{+1.3}$  \\
  $\left\langle r^2\right\rangle _p$ $[fm^2]$ & $0.27_{-0.02}^{+0.01}$  &
$0.40_{-0.04}^{+0.02}$  & $0.37_{-0.03}^{+0.02}$      \\
  $\left\langle r^2\right\rangle _n$ $[fm^2]$ & $-0.061_{-0.007}^{+0.011}$ &
$-0.08_{-0.02}^{+0.03}$  & $-0.07_{-0.02}^{+0.03}$     \\ \hline
  \end{tabular}
  \caption{Results for the polarizabilities and charge radii for different
forms of potentials}
\end{table}
Although the absolute values for the charge radii are still too small they
acquired the right sign. The results for the superposition of the linear \&
Coulombic potential and the Richardson potantial lie within the error bars
of the the experimental results. The magnetic polarizabilities are on the
upper limit and the electric quantities on the lower edge. The magnitutes of
the charge radii are too small by about 50\%. Overall, the ability of the
diquark model to give reasonable values for all the static properties of the
nucleon are very encouraging.

Analyzing the contributions of the different terms shows us that an increase
of the terms containing the spatial wavefunction would lower the electric
and raise the magnetic values of the polarizabilities. A pion cloud around
the nucleon core gives exactly such a contribution. Such sources were
already included by Weiner and Weise\cite{weiner} and had considerable
effects on the electric polarizabilities. Pion clouds will also effect the
charge radii. The transitions $p\rightarrow n+\pi ^{+}$ give positive
contribution and $n\rightarrow p+\pi ^{-}$ yields the desired negative part.
Therefore we have begun to consider the influence the pion cloud will have
on the constraints $\frac{g_a}{g_v}$ and $\frac{\mu _p}{\mu _n}$ and to find
the corresponding values for the mixing $a$. The core values can be directly
taken from this paper - the pion cloud terms will have to be added.

{\bf{Acknowledgements:}}
Partial support for this work was provided by a grant from the
U.S.Department of Energy, DE-FG02-92ER40702.

\newpage
{\bf{Figure Captions}}

1. Electric (a), magnetic (b) polarizabilities and charge radii (c) for
neutron (solid line) and proton (dashed line) as a function of the
mixing parameter $a$, which is explained in the text.

\end{document}